\documentclass[aps,prb,twocolumn,showpacs,floats]{revtex4}
\usepackage{graphics}
\usepackage{amssymb}
\usepackage{amsmath}
\begin{document}

\title{Scattering states of a vortex in the proximity-induced superconducting state at the interface of a topological insulator and an $s$-wave superconductor}
\author{Adam C. Durst}
\affiliation{Department of Physics and Astronomy, Hofstra University, Hempstead, NY 11549-1510}
\date{October 5, 2015}

\begin{abstract}
We consider an isolated vortex in the two-dimensional proximity-induced superconducting state formed at the interface of a three-dimensional strong topological insulator (TI) and an $s$-wave superconductor (sSC).  Prior calculations of the bound states of this system famously revealed a zero-energy state that is its own conjugate, a Majorana fermion bound to the vortex core.  We calculate, not the bound states, but the scattering states of this system, and ask how the spin-momentum-locked massless Dirac form of the single-particle Hamiltonian, inherited from the TI surface, affects the cross section for scattering Bogoliubov quasiparticles from the vortex.  As in the case of an ordinary superconductor, this is a two-channel problem with the vortex mixing particle-like and hole-like excitations.  And as in the ordinary case, the same-channel differential cross section diverges in the forward direction due to the Aharonov-Bohm effect, resulting in an infinite total cross section but finite transport and skew cross sections.  We calculate the transport and skew cross sections numerically, via a partial wave analysis, as a function of both quasiparticle excitation energy and chemical potential.  Novel effects emerge as particle-like or hole-like excitations are tuned through the Dirac point.
\end{abstract}


\maketitle

\section{Introduction}
\label{sec:intro}
In 1964, Caroli, de Gennes, and Matricon \cite{car64,deg99} calculated the bound state spectrum of a vortex in a type-II $s$-wave superconductor, the well-known Caroli-de Gennes-Matricon states.  In 1968, Cleary \cite{cle68,cle70} expanded on this work by calculating the scattering states.  He considered the problem of a single vortex and used a partial wave analysis to compute the cross section for quasiparticles scattering therefrom.  In 2008, Fu and Kane \cite{fu08} considered a new type of superconductor \cite{sac11,vel12}, the proximity-induced superconducting state at the interface of a topological insulator (TI) \cite{has10,qi11,moo10,kan05a,kan05b,fu07a,moo07,roy09,ber06,kon07,fu07b,hsi08,zha09,liu10,xia09} and an $s$-wave superconductor (sSC).  Since the excitations of the TI surface are spin-polarized massless Dirac fermions, proximity to the $s$-wave superconductor yields an exotic superconducting state at the TI-sSC interface, with novel quasiparticle excitations.  Applying the Caroli-de Gennes-Matricon analysis to this new superconductor, Fu and Kane \cite{fu08} calculated the bound state spectrum of a single vortex and showed that there exists a zero energy bound state that is its own conjugate, an example of a Majorana fermion \cite{maj37,lee14,ali12,bee13,law09,sau10,hos11,akz14}.

In this paper, we consider the scattering states, playing Cleary to Fu and Kane's Caroli-de Gennes-Matricon.  We ask how quasiparticles with excitation energy greater than the superconducting gap, $E>\Delta_0$, scatter from a vortex in this novel superconducting state at the TI-sSC interface.  Employing a partial wave approach similar to Cleary's analysis for an ordinary superconductor \cite{cle68}, we calculate the cross section for this topological case.  We have used such an approach previously, for the case of an ordinary (non-topological) $d$-wave superconductor, in Refs.~\onlinecite{dur03,kul11,gan11}.  That $d$-wave case involved the extra complication of linearizing about an off-origin nodal point, which resulted in a non-central effective potential that mixed angular momentum eigenstates.  The current $s$-wave topological case is simpler in the sense that the Bogoliubov-de Gennes (BdG) Hamiltonian \cite{deg99} commutes with the total angular momentum operator, such that each angular-momentum-indexed eigenstate can be calculated independently.  Spin-up and spin-down are mixed by the massless Dirac Hamiltonian inherited from the TI surface, and particle and hole are mixed by the superconducting order parameter in the BdG equation, but radial functions for different angular momenta are not mixed by the vortex.  Thus, ours is a $4 \times 4$ matrix problem in spin-particle-hole space for each angular-momentum-indexed eigenstate.  Solution of the resulting fourth order differential equations and application of proper boundary conditions yields the scattering cross sections that we seek.

We begin in Sec.~\ref{sec:formulation} by writing down the Dirac-Bogoliubov-de Gennes equation, separating variables in polar coordinates, and defining a hard-cutoff model for the vortex.  In Sec.~\ref{sec:radial}, we solve the resulting radial equations in each of the three regions defined by our model: inside the vortex core, inside the vortex but outside the core, and outside the vortex.  In Sec.~\ref{sec:crosssection}, we consider the asymptotic form of our eigenstates and use them to construct incident plane waves and outgoing radial waves, thereby defining scattering amplitudes in terms of a set of angular-momentum-indexed coefficients.  By matching solutions at region boundaries and imposing appropriate restrictions at small and large distances, we develop an algorithm for obtaining these coefficients and therefore the scattering cross sections.  Numerical results obtained by applying these algorithms for parameter values of interest are presented in Sec.~\ref{sec:results}. Conclusions are discussed in Sec.~\ref{sec:conclusions}.

\section{Formulation}
\label{sec:formulation}

\subsection{Dirac-Bogoliubov-de Gennes equation}
\label{ssec:DBdG}
We seek the scattering states for a vortex in the proximity-induced 2D superconducting state at the interface of a strong 3D topological insulator (TI) and a conventional $s$-wave superconductor (sSC).  Our approach is to solve the Bogoliubov-de Gennes equation \cite{deg99}
\begin{equation}
H \Phi = E \Phi
\label{eq:DBdG}
\end{equation}
\begin{equation}
H = \left[ \begin{array}{cc} H_0({\bf A})-\mu & \Delta(r)e^{i\theta} \\ \Delta(r) e^{-i\theta} & \mu-H_0(-{\bf A}) \end{array} \right]
\label{eq:Hdef}
\end{equation}
in the presence of a single vortex that twists the phase of the order parameter once around the origin, depletes its magnitude $\Delta(r)$, and contributes the vector potential ${\bf A}$.  For the case at hand, this is sometimes referred to as the Dirac-Bogoliubov-de Gennes (DBdG) equation \cite{bee08,bee06} since the single-particle Hamiltonian for the TI surface, $H_0$, has the massless Dirac form
\begin{equation}
H_0({\bf A}) = v \vec{\sigma} \cdot \left( -i\vec{\nabla} - e{\bf A}/c \right)
\label{eq:H0def}
\end{equation}
where $v$ is the slope of the Dirac cone and $\vec{\bf \sigma}=(\sigma_1, \sigma_2)$ are spin-space Pauli matrices.  In the above, $\mu$ is the chemical potential, we have adopted units where $\hbar=1$, and $\Phi(r,\theta)$ is the four-component (spin-up-particle, spin-down-particle, spin-up-hole, spin-down-hole) wave function.  Note that the TI surface Hamiltonian mixes spin-up and spin-down and the order parameter term mixes particle and hole.  Therefore the eigenstates of the Hamiltonian will be superpositions of all four components, with neither spin nor charge well defined.  Note also that $\Delta$ is a proximity-induced order parameter \cite{vol95}, dependent upon the transmittance of the TI-sSC interface, and of a form inherited from that of the $s$-wave superconductor \cite{fu08,bee06}.

At this stage, it is convenient to apply the unitary transformation
\begin{equation}
U = \left[ \begin{array}{cc} e^{-i\frac{\theta}{2}} & 0 \\ 0 & e^{i\frac{\theta}{2}} \end{array} \right]
\label{eq:Udef}
\end{equation}
which strips off the phase of the order parameter at the expense of imposing antiperiodic boundary conditions on the transformed wave function.
\begin{equation}
\Psi \equiv U \Phi \;\;\;\;\;\; \Psi(\theta + 2\pi) = - \Psi(\theta)
\label{eq:Psidef}
\end{equation}
This antiperiodic wave function satisfies $H^\prime\Psi = E\Psi$
\begin{equation}
H^\prime \equiv UHU^{-1} = -iv\tau_3\vec{\sigma}\cdot\vec{\nabla} - \mu \tau_3 + v\vec{\sigma}\cdot{\bf P}_s + \Delta\tau_1
\label{eq:Hprimedef}
\end{equation}
where we have introduced a second set of Pauli matrices $\{\tau_1, \tau_2, \tau_3\}$ that act upon particle-hole space, and have defined the superfluid momentum function
\begin{equation}
{\bf P}_s \equiv \frac{1}{2}{\bf \nabla}\theta - \frac{e}{c}{\bf A} = P_s(r)\hat{\theta} .
\label{eq:Psdef}
\end{equation}

\subsection{Separation of variables in polar coordinates}
\label{ssec:sepvar}

It is straightforward to show that $H^\prime$ commutes with the total angular momentum operator
\begin{equation}
J \equiv -i\frac{\partial}{\partial\theta} + \frac{\sigma_3}{2}
\label{Jdef}
\end{equation}
so there exists a complete set of simultaneous eigenstates
\begin{equation}
J \Psi_n = n \Psi_n
\label{eq:Jeig}
\end{equation}
\begin{equation}
H^\prime \Psi_n = E \Psi_n .
\label{eq:Eeig}
\end{equation}
Note that Eq.~(\ref{eq:Jeig}) is satisfied by any solution of the form
\begin{equation}
\Psi_n(r,\theta) = \left[ \begin{array}{c} f_{n\uparrow}(r) e^{i(n-\frac{1}{2})\theta} \\ f_{n\downarrow}(r) e^{i(n+\frac{1}{2})\theta} \\
g_{n\uparrow}(r) e^{i(n-\frac{1}{2})\theta} \\ g_{n\downarrow}(r) e^{i(n+\frac{1}{2})\theta} \end{array} \right]
\label{eq:Psindef}
\end{equation}
where $f_{n\uparrow}$, $f_{n\downarrow}$, $g_{n\uparrow}$, and $g_{n\downarrow}$ are as-yet-undermined scalar radial functions and the antiperiodic boundary conditions require $n$ to be an integer.  Plugging a solution of this form into Eq.~(\ref{eq:Eeig}) reduces it to a set of four coupled first-order ordinary differential equations for the four radial functions.  In matrix form, these radial equations become
\begin{equation}
\frac{\partial {\bf z}}{\partial \rho} = {\bf M}(\rho) {\bf z}
\label{eq:matrixradial}
\end{equation}
\begin{equation}
{\bf M}(\rho) =
\left[ \begin{array}{cccc} \frac{n-\frac{1}{2}+\rho \tilde{P}_s}{\rho} & i\alpha & 0 & -i\tilde{\Delta} \\
i\alpha & -\frac{n+\frac{1}{2}+\rho \tilde{P}_s}{\rho} & -i\tilde{\Delta} & 0 \\
0 & i\tilde{\Delta} & \frac{n-\frac{1}{2}-\rho \tilde{P}_s}{\rho} & -i\beta \\
i\tilde{\Delta} & 0 & -i\beta & -\frac{n+\frac{1}{2}-\rho \tilde{P}_s}{\rho} \end{array} \right]
\label{eq:Mmatrix}
\end{equation}
where ${\bf z} \equiv [f_{n\uparrow}, f_{n\downarrow}, g_{n\uparrow}, g_{n\downarrow}]^T$ and we have defined dimensionless parameters by writing all energies in ratio with the gap maximum $\Delta_0$ and all lengths in product with $k_0 \equiv \Delta_0/v$:
\begin{equation}
\rho \equiv k_0 r \;\;\;\;\;\; \tilde{\Delta}(\rho) \equiv \frac{\Delta}{\Delta_0} \;\;\;\;\;\; \tilde{P}_s(\rho) \equiv \frac{P_s}{k_0}
\label{eq:dim1}
\end{equation}
\begin{equation}
\alpha \equiv \frac{E+\mu}{\Delta_0} \;\;\;\;\;\; \beta \equiv \frac{E-\mu}{\Delta_0}
\label{eq:dim2}
\end{equation}
Our task moving forward is to solve these radial equations (see Sec.~\ref{sec:radial}) and use the resulting eigenstates to construct the scattering cross section (see Sec.~\ref{sec:crosssection}).

\subsection{Model}
\label{ssec:model}
It is clear from the above that our results will depend on the excitation energy of the incident quasiparticle, $E$, and the chemical potential, $\mu$, as well as the radial dependence of the superflow and gap functions, $\tilde{P}_s(\rho)$ and $\tilde{\Delta}(\rho)$, in the vicinity of the vortex.  As per Eq.~(\ref{eq:Psdef}), and noting that ${\bf \nabla}\theta=\hat{\theta}/r$, the superflow function initially falls off like $1/2\rho$ and then is fully suppressed at radial distances large compared with the London penetration depth, $\lambda$.  The gap function is suppressed at distances small compared with the superconducting coherence length, $\xi$, and then restored to $\Delta_0$ at larger distances.  While the detailed form of these functions can be complex and material-dependent, we attempt to capture the essential physics within a tractable calculation by adopting a simple model with hard cutoffs at the two length scales:
\begin{equation}
\tilde{P}_s(\rho) = \left\{ \begin{array}{r} 1/2\rho \;\;\; \mbox{for} \;\;\; \rho < k_0 \lambda \\ 0 \;\;\;\;\;\; \mbox{for} \;\;\; \rho > k_0 \lambda \end{array} \right.
\label{eq:Psmodel}
\end{equation}
\begin{equation}
\tilde{\Delta}(\rho) = \left\{ \begin{array}{r} 0 \;\;\; \mbox{for} \;\;\; \rho < k_0 \xi \\ 1 \;\;\; \mbox{for} \;\;\; \rho > k_0 \xi \end{array} \right.
\label{eq:Deltamodel}
\end{equation}
For the type-II ($\lambda > \xi$) case that we consider, this model defines three distinct regions: (1) Inside the vortex core ($r < \xi$), where the order parameter vanishes but there is nonzero superflow, (2) Inside the vortex but outside the core ($\xi < r < \lambda$), where both the order parameter and the superflow are nonzero, and (3) Outside the vortex ($r > \lambda$), where the order parameter is nonzero but the superflow vanishes.  In the following section, we solve the radial equations in each of these regions, with solutions to be matched at the boundaries.

\section{Solution of Radial Equations}
\label{sec:radial}

\subsection{Inside vortex core}
\label{ssec:insidecore}
Inside the vortex core ($r < \xi$), $\tilde{P}_s(\rho) = 1/2\rho$ and $\tilde{\Delta}=0$.  Without an order parameter to couple particle and hole, the two sectors decouple and the ${\bf M}$-matrix of Eq.~(\ref{eq:Mmatrix}) becomes block diagonal.
\begin{equation}
{\bf M}(\rho) =
\left[ \begin{array}{cccc} \frac{n}{\rho} & i\alpha & 0 & 0 \\
i\alpha & -\frac{n+1}{\rho} & 0 & 0 \\
0 & 0 & \frac{n-1}{\rho} & -i\beta \\
0 & 0 & -i\beta & -\frac{n}{\rho} \end{array} \right]
\label{eq:Minsidecore}
\end{equation}
The particle sector of Eq.~(\ref{eq:matrixradial}) reduces to two coupled first order equations
\begin{equation}
\left( \frac{\partial}{\partial\rho} - \frac{n}{\rho} \right) f_{n\uparrow} = i\alpha f_{n\downarrow}
\label{eq:feq1}
\end{equation}
\begin{equation}
\left( \frac{\partial}{\partial\rho} + \frac{n+1}{\rho} \right) f_{n\downarrow} = i\alpha f_{n\uparrow}
\label{eq:feq2}
\end{equation}
which combine to yield
\begin{equation}
\left( \frac{\partial^2}{\partial\rho^2} + \frac{1}{\rho} \frac{\partial}{\partial\rho} + \alpha^2 - \frac{n^2}{\rho^2} \right) f_{n\uparrow} = 0
\label{eq:fbesseleq}
\end{equation}
which is just the Bessel equation for variable $|\alpha|\rho$ and index $n$.  Solutions for $f_{n\uparrow}$ are therefore integer Bessel functions of order $n$ and argument $|\alpha|\rho$
\begin{equation}
f_{n\uparrow}(\rho) = A_n^1 J_n(|\alpha|\rho) + A_n^2 Y_n(|\alpha|\rho)
\label{eq:fnup}
\end{equation}
where $A_n^1$ and $A_n^2$ are complex constants to be fixed by boundary conditions.  Substituting back into Eq.~(\ref{eq:feq1}) and making use of the Bessel function identity \cite{gra94} $(\partial/\partial x - \nu/x) Z_\nu(x) = -Z_{\nu+1}(x)$ reveals that
\begin{equation}
f_{n\downarrow}(\rho) = iA_n^1 J_{n+1}(|\alpha|\rho) + iA_n^2 Y_{n+1}(|\alpha|\rho)
\label{eq:fndown}
\end{equation}
where the constants are the same as above.  Similarly, the hole sector of Eq.~(\ref{eq:matrixradial}) reduces to
\begin{equation}
\left( \frac{\partial}{\partial\rho} - \frac{n-1}{\rho} \right) g_{n\uparrow} = -i\beta g_{n\downarrow}
\label{eq:geq1}
\end{equation}
\begin{equation}
\left( \frac{\partial}{\partial\rho} + \frac{n}{\rho} \right) g_{n\downarrow} = -i\beta g_{n\uparrow}
\label{eq:geq2}
\end{equation}
Solutions are once again integer Bessel functions, this time of argument $|\beta|\rho$.  We therefore find that
\begin{equation}
g_{n\uparrow}(\rho) = A_n^3 J_{n-1}(|\beta|\rho) + A_n^4 Y_{n-1}(|\beta|\rho)
\label{eq:gnup}
\end{equation}
\begin{equation}
g_{n\downarrow}(\rho) = -iA_n^3 J_n(|\beta|\rho) - iA_n^4 Y_n(|\beta|\rho)
\label{eq:gndown}
\end{equation}
where $A_n^3$ and $A_n^4$ are additional complex constants.  Since solutions inside the vortex core must be well-behaved at the origin, the coefficients multiplying the Y-Bessel functions must be set to zero.
\begin{equation}
A_n^2 = A_n^4 = 0
\label{eq:A2A4zero}
\end{equation}
Finally, combining all four components, we can write down the vector solution of Eq.~(\ref{eq:matrixradial}) inside the vortex core:
\begin{equation}
{\bf x}_n(\rho) = A_n^1 {\bf x}_n^1(\rho) + A_n^3 {\bf x}_n^3(\rho)
\label{eq:xn}
\end{equation}
\begin{equation}
{\bf x}_n^1(\rho) = \left[ \begin{array}{c} J_n(|\alpha|\rho) \\ iJ_{n+1}(|\alpha|\rho) \\ 0 \\ 0 \end{array} \right]
\;\;\;\;\;\;
{\bf x}_n^3(\rho) = \left[ \begin{array}{c} 0 \\ 0 \\ J_{n-1}(|\beta|\rho) \\ -iJ_n(|\beta|\rho) \end{array} \right]
\label{eq:xn1xn3}
\end{equation}

\subsection{Inside vortex, outside core}
\label{ssec:insidevortex}
Outside the vortex core, but still inside the vortex ($\xi < r < \lambda$), $\tilde{P}_s(\rho) = 1/2\rho$ and $\tilde{\Delta}=1$.  Thus, the particle and hole sectors are coupled by the order parameter, and the ${\bf M}$-matrix is no longer block diagonal.
\begin{equation}
{\bf M}(\rho) =
\left[ \begin{array}{cccc} \frac{n}{\rho} & i\alpha & 0 & -i \\
i\alpha & -\frac{n+1}{\rho} & -i & 0 \\
0 & i & \frac{n-1}{\rho} & -i\beta \\
i & 0 & -i\beta & -\frac{n}{\rho} \end{array} \right]
\label{eq:Minsidevortex}
\end{equation}
In this region, the radial functions are most simply attained through direct integration of Eq.~(\ref{eq:matrixradial}).  We posit a solution of the form
\begin{equation}
{\bf y}_n(\rho) = B_n^1 {\bf y}_n^1(\rho) + B_n^2 {\bf y}_n^2(\rho) + B_n^3 {\bf y}_n^3(\rho) + B_n^4 {\bf y}_n^4(\rho)
\label{eq:yn}
\end{equation}
where the $B_n^j$ are as-yet-undetermined complex constants and we define four linearly-independent initial conditions at $\rho = k_0 \xi$
\begin{equation}
{\bf y}_n^j(\rho=k_0\xi) = \left\{
\left[ \begin{array}{c} 1 \\ 0 \\ 0 \\ 0 \end{array} \right],
\left[ \begin{array}{c} 0 \\ 1 \\ 0 \\ 0 \end{array} \right],
\left[ \begin{array}{c} 0 \\ 0 \\ 1 \\ 0 \end{array} \right],
\left[ \begin{array}{c} 0 \\ 0 \\ 0 \\ 1 \end{array} \right] \right\}
\label{eq:ynjinitial}
\end{equation}
We then use an RK4 Runge-Kutta algorithm to numerically propagate each of these solutions out to $\rho = k_0 \lambda$.  Continuity at both boundaries will later be used to determine the constants.

\subsection{Outside vortex}
\label{ssec:outsidevortex}
Outside the vortex ($r > \lambda$), $\tilde{P}_s(\rho) = 0$ and $\tilde{\Delta}=1$.  This is the bulk of the superconductor, so the order parameter still couples the particle and hole sectors, but there is no superflow.  Thus, the radial equations take the form
\begin{equation}
\frac{\partial {\bf z}}{\partial \rho} = \left( \frac{\bf B}{\rho} + {\bf A}_0 \right) {\bf z}
\label{eq:matrixradialoutside}
\end{equation}
\begin{equation}
{\bf B} =
\left[ \begin{array}{cccc} n-\frac{1}{2} & 0 & 0 & 0 \\
0 & -(n+\frac{1}{2}) & 0 & 0 \\
0 & 0 & n-\frac{1}{2} & 0 \\
0 & 0 & 0 & -(n+\frac{1}{2}) \end{array} \right]
\label{eq:Bdef}
\end{equation}
\begin{equation}
{\bf A}_0 = i
\left[ \begin{array}{cccc} 0 & \alpha & 0 & -1 \\
\alpha & 0 & -1 & 0 \\
0 & 1 & 0 & -\beta \\
1 & 0 & -\beta & 0 \end{array} \right]
\label{eq:A0def}
\end{equation}
In this region, we seek solutions which, in the asymptotic limit ($\rho \rightarrow \infty$), can be used to construct incident plane waves and outgoing radial waves and thereby obtain scattering amplitudes.  We therefore solve Eq.~(\ref{eq:matrixradialoutside}) via asymptotic series expansion \cite{ben99} about the irregular singular point at $\rho \rightarrow \infty$.

Note that were it not for the particle-hole coupling, the solutions for each sector would be half-integer Bessel functions.  Thus, it makes sense to try vector solutions with the same $e^{i\lambda_j\rho}/\sqrt{\rho}$ leading behavior, expanding in powers of $1/\rho$.  Labeling each of the four such solutions by $j=\{1,2,3,4\}$, we try
\begin{equation}
{\bf z}^j(\rho) = \frac{e^{i\lambda_j\rho}}{\sqrt{\rho}} \sum_{m=0}^\infty {\bf a}_m^j \rho^{-m}
\label{eq:zj}
\end{equation}
where the $\lambda_j$ and ${\bf a}_m^j$ are constant scalars and constant vectors respectively.  Plugging into Eq.~(\ref{eq:matrixradialoutside}), we obtain the following vector recursion relations for $m=0$
\begin{equation}
\left( {\bf A}_0 - i\lambda_j \right) {\bf a}_0^j = 0
\label{eq:eigen}
\end{equation}
and for $m=1,2,3,...$
\begin{equation}
\left( {\bf A}_0 - i\lambda_j \right) {\bf a}_m^j = -\left( {\bf B} + m - \frac{1}{2} \right) {\bf a}_{m-1}^j .
\label{eq:recursion}
\end{equation}
Note that Eq.~(\ref{eq:eigen}) is just the eigenvalue equation for the matrix ${\bf A}_0$, which requires that $i\lambda_j$ and ${\bf a}_0^j$ are the eigenvalues and eigenvectors of ${\bf A}_0$.  Solving this eigensystem reveals that
\begin{eqnarray}
\lambda_1 &=& \sqrt{\big(\tfrac{E}{\Delta_0}\big)^2 - 1} + \tfrac{\mu}{\Delta_0} \nonumber \\
\lambda_2 &=& \sqrt{\big(\tfrac{E}{\Delta_0}\big)^2 - 1} - \tfrac{\mu}{\Delta_0} \nonumber \\
\lambda_3 &=& -\lambda_1 \nonumber \\
\lambda_4 &=& -\lambda_2
\label{eq:eigenvalues}
\end{eqnarray}
and
\begin{equation}
{\bf a}_0^1 = \left[ \begin{array}{r} a \\ a \\ b \\ b \end{array} \right]
\;\;\;\;
{\bf a}_0^2 = \left[ \begin{array}{r} b \\ -b \\ a \\ -a \end{array} \right]
\;\;\;\;
{\bf a}_0^3 = \left[ \begin{array}{r} a \\ -a \\ b \\ -b \end{array} \right]
\;\;\;\;
{\bf a}_0^4 = \left[ \begin{array}{r} b \\ b \\ a \\ a \end{array} \right]
\label{eq:eigenvectors}
\end{equation}
where
\begin{equation}
a = \frac{1}{\sqrt{2[1 + (\alpha-\lambda_1)^2]}}
\;\;\;\;\;\;
b = (\alpha - \lambda_1) a .
\label{eq:abdef}
\end{equation}
Though Eq.~(\ref{eq:recursion}) has the form of a recursion relation, its utility in obtaining the rest of the vector coefficients, ${\bf a}_m^j$, is considerably more subtle than for a typical (scalar) recursion relation because it is a matrix relation between vectors.  To illustrate this subtlety, let's define ${\bf A}^\prime \equiv {\bf A}_0 - i\lambda_j$ and ${\bf B}_m^\prime \equiv -({\bf B} + m -1/2)$ such that Eqs.~(\ref{eq:eigen}) and (\ref{eq:recursion}) can be written as
\begin{equation}
{\bf A}^\prime {\bf a}_0 = 0
\label{eq:eig}
\end{equation}
\begin{equation}
{\bf A}^\prime {\bf a}_m = {\bf B}_m^\prime {\bf a}_{m-1}
\label{eq:recur}
\end{equation}
where we have suppressed the $j$ superscripts.  It is clear from Eq.~(\ref{eq:eig}) that ${\bf A}^\prime$ is singular, with a null space (one-dimensional, as long as the $\lambda_j$ are distinct) defined by ${\bf a}_0$.  Thus, Eq.~(\ref{eq:recur}) can only be satisfied if ${\bf B}_m^\prime {\bf a}_{m-1}$ lies in the (typically three-dimensional) column space of ${\bf A}^\prime$ and is therefore orthogonal to the left null space of ${\bf A}^\prime$, defined by the vector ${\bf b}_0$ where
\begin{equation}
{\bf A}^{\prime\dagger} {\bf b}_0 = 0 .
\label{eq:leftnulldef}
\end{equation}
So in addition to constraining ${\bf a}_m$, Eq.~(\ref{eq:recur}) constrains ${\bf a}_{m-1}$ by requiring that
\begin{equation}
{\bf b}_0^\dagger {\bf B}_m^\prime {\bf a}_{m-1} = 0 .
\label{eq:perptoleftnull}
\end{equation}

This is particularly worrisome for $m=1$, since we have no room to adjust ${\bf a}_0$ as it is fully specified (up to a multiplicative factor) by Eq.~(\ref{eq:eig}) and defined explicitly in Eq.~(\ref{eq:eigenvectors}).  In order for a solution of the desired form to exist, it must be the case that the eigenvectors, ${\bf a}_0^j$, satisfy
\begin{equation}
{\bf b}_0^\dagger {\bf B}_1^\prime {\bf a}_0 = 0 .
\label{eq:a0condition}
\end{equation}
Fortunately, they do.  We can see this by solving Eq.~(\ref{eq:leftnulldef}) to find the left null space \cite{str88} vectors for $j=\{1,2,3,4\}$.  In each case we see that ${\bf b}_0 = \tau_3 {\bf a}_0$.  Noting that ${\bf B}_1^\prime = -n\sigma_3$, writing the components of ${\bf a}_0$ as $a_{0i}$, and plugging into Eq.~(\ref{eq:a0condition}), we find that
\begin{equation}
{\bf b}_0^\dagger {\bf B}_1^\prime {\bf a}_0 = -n {\bf a}_0^\dagger \tau_3 \sigma_3 {\bf a}_0 = -n\left(a_{01}^2 - a_{02}^2 - a_{03}^2 + a_{04}^2 \right)
\label{eq:a0conditionsatisfied}
\end{equation}
which is zero since $a_{01} = \pm a_{02}$ and $a_{03} = \pm a_{04}$ for each of the eigenvectors in Eq.~(\ref{eq:eigenvectors}).

Having checked this, we can obtain ${\bf a}_1$ from ${\bf a}_0$ via Eq.~(\ref{eq:recur}).  Since ${\bf B}_1^\prime {\bf a}_0$ is in the column space of ${\bf A}^\prime$, there exists an ${\bf a}_1$ that satisfies the recursion relation.  In fact, there are many, since we can always add to it a vector in the null space of ${\bf A}^\prime$ to generate another.  Multiplying by the Moore-Penrose pseudoinverse \cite{str88}, ${\bf A}_{PI}^\prime$, obtained via singular value decomposition of ${\bf A}^\prime$, provides the smallest such vector, since it lacks a null space component.  But this freedom to add a null space component is precisely what we need in order to ensure that ${\bf a}_1$ satisfies Eq.~(\ref{eq:perptoleftnull}) for $m=2$.  For general $m=1,2,3,...$, given a vector ${\bf a}_{m-1}$ that satisfies Eq.~(\ref{eq:perptoleftnull}), our algorithm to find ${\bf a}_m$ is as follows:  (1) Use the pseudoinverse to calculate the row space component.
\begin{equation}
{\bf a}_m^{\rm row} = {\bf A}_{PI}^\prime {\bf B}_m^\prime {\bf a}_{m-1}
\label{eq:arow}
\end{equation}
(2) Apply the condition in Eq.~(\ref{eq:perptoleftnull}) for $m \rightarrow m+1$ and compute the extent to which it fails.
\begin{equation}
\gamma_{LN} \equiv {\bf b}_0^\dagger {\bf B}_{m+1}^\prime {\bf a}_m^{\rm row}
\label{eq:leftnullcomponent}
\end{equation}
This is the component of the right-hand-side of the next iteration of the recursion relation that does not lie within the column space of ${\bf A}^\prime$.  (3) Add just enough of the null space vector, ${\bf a}_0$, to cancel this component.
\begin{equation}
{\bf a}_m = {\bf a}_m^{\rm row} + {\bf a}_m^{\rm null}
\;\;\;\;\;\;
{\bf a}_m^{\rm null} \equiv -\frac{\gamma_{LN}}{{\bf b}_0^\dagger {\bf B}_{m+1}^\prime {\bf a}_0} {\bf a}_0
\label{eq:anull}
\end{equation}
The result is the vector
\begin{equation}
{\bf a}_m = {\bf A}_{PI}^\prime {\bf B}_m^\prime {\bf a}_{m-1}
- \frac{{\bf b}_0^\dagger {\bf B}_{m+1}^\prime {\bf A}_{PI}^\prime {\bf B}_m^\prime {\bf a}_{m-1}}{{\bf b}_0^\dagger {\bf B}_{m+1}^\prime {\bf a}_0} {\bf a}_0
\label{eq:atot}
\end{equation}
which satisfies Eq.~(\ref{eq:recur}) for the current iteration as well as Eq.~(\ref{eq:perptoleftnull}) for the next one.  Repeating this algorithm for each successive iteration, and for each of the four $j$'s, yields the vector coefficients in the asymptotic series solutions defined in Eq.~(\ref{eq:zj}).

When computing the asymptotic series solution for the simpler case of the half-integer Bessel functions \cite{ben99}, one finds that the series actually terminates at a maximum value of $m$.  It turns out that (at least within the numerical precision of our computation), our asymptotic series terminates as well, with $m_{\rm max} = |n|$, where $n$ is the eigenvalue of total angular momentum defined in Eq.~(\ref{eq:Jeig}).  This is quite fortunate and perhaps, in hindsight, not altogether surprising, as our problem can be thought of as two coupled half-integer Bessel equations, one for the particle sector and one for the hole sector. Thus, within numerical precision, these asymptotic series solutions are {\it exact}, valid everywhere outside the vortex.  A linear combination of our four series solutions yields the general solution to the radial equations in the region outside the vortex ($\rho > k_0\lambda$).  Hence, we write
\begin{equation}
{\bf z}_n(\rho) = C_n^1 {\bf z}_n^1(\rho) + C_n^2 {\bf z}_n^2(\rho) + C_n^3 {\bf z}_n^3(\rho) + C_n^4 {\bf z}_n^4(\rho)
\label{eq:zn}
\end{equation}
where the $C_n^j$ are complex constants (to be determined by matching boundary conditions) and each of the four {\it finite} asymptotic series solutions are expressed via
\begin{equation}
{\bf z}_n^j(\rho) = \frac{e^{i\lambda_j\rho}}{\sqrt{\rho}} \sum_{m=0}^{|n|} {\bf a}_m^j \rho^{-m}
\label{eq:zjfinite}
\end{equation}
with eigenvalues $\lambda_j$ and vector coefficients ${\bf a}_m^j$ determined via the algorithm above.

\section{Constructing the cross section}
\label{sec:crosssection}

\subsection{Asymptotic wave function}
\label{ssec:asymptoticWF}
With the radial functions in hand, our next step is to consider the asymptotic form of the wave function and use it to construct an incident plane wave and outgoing radial wave, from which we can extract the scattering amplitudes and cross sections.  In the asymptotic ($\rho \rightarrow \infty$) limit, we use the outside-the-vortex radial functions (Eq.~(\ref{eq:zn})) and need only retain the leading ($m=0$) behavior.
\begin{equation}
{\bf z}_n(\rho) = \left[ \begin{array}{c} f_{n\uparrow} \\ f_{n\downarrow} \\ g_{n\uparrow} \\ g_{n\downarrow} \end{array} \right]
= \sum_{j=1}^4 C_n^j \frac{e^{i\lambda_j \rho}}{\sqrt{\rho}} {\bf a}_0^j
\label{eq:znasymptotic}
\end{equation}
The transformed ($2\pi$-antiperiodic) eigenstate, $\Psi_n(\rho,\theta)$, is obtained by plugging the radial functions into Eq.~(\ref{eq:Psindef}), which can then be multiplied by the inverse transformation matrix, $U^{-1}=e^{i\tau_3 \theta/2}$, to yield the $2\pi$-periodic eigenstate
\begin{equation}
\Phi_n = e^{in\theta}
\left[ \begin{array}{l} f_{n\uparrow} \\ f_{n\downarrow} e^{i\theta} \\ g_{n\uparrow} e^{-i\theta} \\ g_{n\downarrow} \end{array} \right]
= e^{in\theta} \sum_{j=1}^4 C_n^j \frac{e^{i\lambda_j \rho}}{\sqrt{\rho}}
\left[ \begin{array}{l} a_{01}^j \\ a_{02}^j e^{i\theta} \\ a_{03}^j e^{-i\theta} \\ a_{04}^j \end{array} \right]
\label{eq:PhinAsymptotic}
\end{equation}
where the $a_{0i}^j$ are the components of the ${\bf a}_0^j$ vectors.  The full asymptotic wave function is then given by a sum over these angular momentum eigenstates.  Summing over $n$ and writing out the four terms explicitly, we obtain
\begin{eqnarray}
\Phi = \!\!\!\!\!\! && \sum_{n=-\infty}^\infty e^{in\theta} \left\{
C_n^1 \frac{e^{i\lambda_1 \rho}}{\sqrt{\rho}} \left[ \begin{array}{l} a \\ a\ e^{i\theta} \\ b\ e^{-i\theta} \\ b \end{array} \right]
+ C_n^2 \frac{e^{i\lambda_2 \rho}}{\sqrt{\rho}} \left[ \begin{array}{l} b \\ -b\ e^{i\theta} \\ a\ e^{-i\theta} \\ -a \end{array} \right]
\right. \nonumber \\
&& + \left. C_n^3 \frac{e^{-i\lambda_1 \rho}}{\sqrt{\rho}} \left[ \begin{array}{l} a \\ -a\ e^{i\theta} \\ b\ e^{-i\theta} \\ -b \end{array} \right]
+ C_n^4 \frac{e^{-i\lambda_2 \rho}}{\sqrt{\rho}} \left[ \begin{array}{l} b \\ b\ e^{i\theta} \\ a\ e^{-i\theta} \\ a \end{array} \right]
\right\}
\label{eq:PhiAsymptotic}
\end{eqnarray}
where the infinite sum is, in practice, truncated at some $n_{\rm max}$ beyond which additional contributions are small.  Typically, quasiparticles of larger incident energy require contributions from more angular momentum eigenstates.

\subsection{Current density functional}
\label{ssec:current}
Setting up the scattering problem requires a definition of the incident and outgoing quasiparticle current, which is provided by the quasiparticle current density functional, ${\bf j}_{qp}[\Phi]$.  This is obtained via continuity
\begin{equation}
\frac{\partial \rho_{qp}}{\partial t} + {\bf \nabla} \cdot {\bf j}_{qp} = 0
\label{eq:continuity}
\end{equation}
with the particle density
\begin{equation}
\rho_{qp} \equiv \Phi^\dagger \Phi = [U^{-1}\Psi]^\dagger [U^{-1}\Psi] = \Psi^\dagger U U^\dagger \Psi = \Psi^\dagger \Psi
\label{eq:density}
\end{equation}
where $\Psi = U \Phi$ is the transformed ($2\pi$-antiperiodic) wave function.  We therefore see that
\begin{equation}
{\bf \nabla} \cdot {\bf j}_{qp} = -\frac{\partial}{\partial t} \left( \Psi^\dagger \Psi \right)
= - \frac{\partial \Psi^\dagger}{\partial t} \psi - \Psi^{\dagger} \frac{\partial \Psi}{\partial t} .
\label{eq:divjqp1}
\end{equation}
The time derivatives are obtained by noting that
\begin{equation}
i\frac{\partial \Psi}{\partial t} = H^\prime \Psi
\label{eq:tdepDBdG}
\end{equation}
where
\begin{equation}
H^\prime = -iv\tau_3\vec{\sigma}\cdot\vec{\nabla} - \mu \tau_3 + \Delta\tau_1
\label{eq:HprimeOutside}
\end{equation}
since ${\bf P}_s = 0$ outside the vortex.  Plugging into Eq.~(\ref{eq:divjqp1}), we find that
\begin{equation}
{\bf \nabla} \cdot {\bf j}_{qp} = {\bf \nabla} \cdot \left( v \Psi^\dagger \vec{\bf \sigma} \tau_3 \Psi \right)
\label{eq:divjqp2}
\end{equation}
and thereby obtain a simple expression for the current density functional
\begin{equation}
{\bf j}_{qp}[\Phi] = v \Phi^\dagger \vec{\bf \sigma} \tau_3 \Phi
\label{eq:current}
\end{equation}
where we have transformed back to the $2\pi$-periodic wave function, $\Phi$, and made use of the fact that $U$ commutes with $\vec{\bf \sigma}\tau_3$.

Written in terms of the four components, $\Phi_i$, of the wave function, this becomes
\begin{equation}
{\bf j}_{qp}[\Phi] = 2v \left( \mbox{Re}\left[ \Phi_1^* \Phi_2 - \Phi_3^* \Phi_4 \right] \hat{\bf x}
+ \mbox{Im}\left[ \Phi_1^* \Phi_2 - \Phi_3^* \Phi_4 \right] \hat{\bf y} \right)
\label{eq:jqpFromComponents}
\end{equation}
So if $\Phi_1^* \Phi_2 - \Phi_3^* \Phi_4 = Q \frac{1}{2} e^{i\phi}$ where $Q$ is a real number, then
\begin{equation}
{\bf j}_{qp}[\Phi] = Q v \left( \cos\phi \hat{\bf x} + \sin\phi \hat{\bf y} \right)
\label{eq:jpqDirection}
\end{equation}
and therefore points in, or opposite to, the direction specified by angle $\phi$, depending on the sign of $Q$.  If $\phi$ is equal to $\theta$, our angular variable, then such a current describes particle flux into (or out of) the origin, in the $\pm \hat{\bf r}$ direction, where
\begin{equation}
\hat{\bf r} \equiv \cos\theta \hat{\bf x} + \sin\theta \hat{\bf y} ,
\label{eq:rhat}
\end{equation}
as is appropriate for an incoming (or outgoing) radial wave.  Whereas, if $\phi$ is equal to $\theta_0$, a constant angle, and $Q$ is positive, then the resulting current describes particle flux in the fixed $\hat{\bf k}$ direction, where
\begin{equation}
\hat{\bf k} \equiv \cos\theta_0 \hat{\bf x} + \sin\theta_0 \hat{\bf y} ,
\label{eq:khat}
\end{equation}
as is appropriate for an incident plane wave.  In the sections that follow, we shall design our incident plane waves and outgoing radial waves such that they produce the correct quasiparticle currents.

\subsection{Incident plane waves}
\label{ssec:incidentPW}
In a normal metal, excitations of a given energy are divided into particle excitations above the Fermi surface and hole excitations below the Fermi surface.  In a superconductor, particles and holes near the Fermi surface are mixed together, but the resulting quasiparticle excitations can still be described as either particle-like or hole-like depending on whether the wavevector yields a single-particle dispersion that is greater than or less than the chemical potential.  In the present case of a proximity-induced superconductor at the TI surface, the single-particle dispersion is that of a massless Dirac cone, with particle-like excitations satifying
\begin{equation}
\pm v k - \mu = +\sqrt{E^2 - \Delta_0^2}
\label{eq:particlelike}
\end{equation}
and hole-like excitations satisfying
\begin{equation}
\pm v k - \mu = -\sqrt{E^2 - \Delta_0^2} .
\label{eq:holelike}
\end{equation}
Comparing these forms with that of the eigenvalues in Eq.~(\ref{eq:eigenvalues}), the physical meaning of those eigenvalues becomes clear.  Eigenvalues $\lambda_1$ and $\lambda_3$ correspond to particle-like excitations, with a wavenumber $k_1 \equiv k_0 \lambda_1$ that satisfies Eq.~(\ref{eq:particlelike}).  And eigenvalues $\lambda_2$ and $\lambda_4$ correspond to hole-like excitations, with a wavenumber $k_2 \equiv k_0 \lambda_2$ that satisfies Eq.~(\ref{eq:holelike}).  Since the vortex couples particle-like and hole-like excitations, the scattering problem we consider is a two-channel problem wherein the scattering event allows for a change of channels.  Incident particle-like excitations yield outgoing particle-like excitations and hole-like excitations, and incident hole-like excitations do the same.  Thus, in what follows, we must consider two cases: that of an incident particle-like plane wave and that of an incident hole-like plane wave.

First, let us consider a particle-like plane wave incident in the $\hat{\bf k} = \cos\theta_0 \hat{\bf x} + \sin\theta_0 \hat{\bf y}$ direction defined by the constant angle $\theta_0$.  Following Ref.~\onlinecite{cle68}, we note that the $2\pi$-antiperiodic representation of the wave function for such such a plane wave has the form
\begin{equation}
\Psi_{PW}^p(\rho,\theta) = e^{-i\frac{\theta-\theta_0}{2}} e^{i{\bf k}_1 \cdot {\bf r}}
\left[ \begin{array}{l} a\ e^{-i\frac{\theta_0}{2}} \\ a\ e^{i\frac{\theta_0}{2}} \\ b\ e^{-i\frac{\theta_0}{2}} \\ b\ e^{i\frac{\theta_0}{2}} \end{array} \right]
\label{eq:PsiPLPW}
\end{equation}
where ${\bf k}_1 \equiv k_0 \lambda_1 \hat{\bf k}$, ${\bf k}_1 \cdot {\bf r} = \lambda_1 \rho \cos(\theta-\theta_0)$, and the component structure of the four-vector is inherited from that of eigenvector ${\bf a}_0^1$.  This form satisfies the DBdG equation outside the vortex, is $2\pi$-antiperiodic in the angular variable $\theta$, is $2\pi$-periodic in the incident angle $\theta_0$, and (as we will show below) yields a current in the $\hat{\bf k}$ direction.  As in the ordinary superconductor case considered by Cleary \cite{cle68}, the half-angle phase twist in the prefactor accounts for the Berry phase (Aharonov-Bohm effect) acquired upon circling the vortex, even at large distances.  Multiplying by $U^{-1}$ yields the $2\pi$-periodic (in $\theta$) representation
\begin{equation}
\Phi_{PW}^p(\rho,\theta) = e^{i{\bf k}_1 \cdot {\bf r}}
\left[ \begin{array}{l} a \\ a\ e^{i\theta_0} \\ b\ e^{-i\theta} \\ b\ e^{-i\theta} e^{i\theta_0} \end{array} \right] .
\label{eq:PhiPLPWdef}
\end{equation}
Looking at the components, we see that
\begin{equation}
\Phi_1^* \Phi_2 - \Phi_3^* \Phi_4 = (a^2 - b^2) e^{i\theta_0}
\label{eq:componentsPLPW}
\end{equation}
so via Eq.~(\ref{eq:jqpFromComponents}), the resulting quasiparticle current is
\begin{equation}
{\bf j}_{qp}[\Phi_{PW}^p] = 2 (a^2 - b^2) v \hat{\bf k} .
\label{eq:currentPLPW}
\end{equation}
A little manipulation of Eqs.~(\ref{eq:eigenvalues}) and (\ref{eq:abdef}) reveals that $a^2 - b^2 > 0$ for all allowed excitation energies ($E>\Delta_0$).  Thus, as per our definition, incident current is in the $+\hat{\bf k}$ direction.

In the other case, that of the hole-like plane wave, also incident in the $\hat{\bf k}$ direction, the wave function is just as above except that ${\bf k}_1$ is replaced by ${\bf k}_2 \equiv k_0 \lambda_2 \hat{\bf k}$ and the component structure of the four-vector is inherited from that of eigenvector ${\bf a}_0^2$ rather than ${\bf a}_0^1$.  In the $2\pi$-periodic representation, this incident hole-like wave function takes the form
\begin{equation}
\Phi_{PW}^h(\rho,\theta) = e^{i{\bf k}_2 \cdot {\bf r}}
\left[ \begin{array}{l} b \\ -b\ e^{i\theta_0} \\ a\ e^{-i\theta} \\ -a\ e^{-i\theta} e^{i\theta_0} \end{array} \right]
\label{eq:PhiHLPWdef}
\end{equation}
which yields precisely the same incident quasiparticle current
\begin{equation}
{\bf j}_{qp}[\Phi_{PW}^h] = 2 (a^2 - b^2) v \hat{\bf k} .
\label{eq:currentHLPW}
\end{equation}

In both cases, it is helpful to make use of the following series expansion of the $e^{i{\bf k}_j \cdot {\bf r}}$ factor:
\begin{eqnarray}
\lefteqn{e^{i{\bf k}_j \cdot {\bf r}} = e^{i\lambda_j \rho \cos(\theta-\theta_0)}} \nonumber \\
&& = \sum_{n=-\infty}^\infty \eta_n i^n J_n(|\lambda_j| \rho) e^{in(\theta-\theta_0)} \nonumber \\
&& = \sum_{n=-\infty}^\infty \frac{e^{in(\theta-\theta_0)}}{\sqrt{2\pi |\lambda_j|}} \left[ e^{-i\frac{\pi}{4}\gamma_j} \frac{e^{i\lambda_j \rho}}{\sqrt{\rho}}
+ (-1)^n e^{i\frac{\pi}{4}\gamma_j} \frac{e^{-i\lambda_j \rho}}{\sqrt{\rho}} \right] \nonumber \\
\label{eq:PWexpansionPos}
\end{eqnarray}
where we have expanded in integer Bessel functions, taken the asymptotic limit, and defined $\gamma_j \equiv \mbox{sgn}(\lambda_j)$ and $\eta_n \equiv \mbox{sgn}(n)^n$.  If we plug the series into Eqs.~(\ref{eq:PhiPLPWdef}) and (\ref{eq:PhiHLPWdef}), precisely as written for components 1 and 3, but with $n$ shifted to $n+1$ for components 2 and 4, we find that the particle-like plane wave takes the form
\begin{eqnarray}
\Phi_{PW}^p(\rho,\theta) && \!\!\!\!\! = \! \sum_{n=-\infty}^\infty e^{in\theta} \left\{ \frac{ e^{-i\frac{\pi}{4}\gamma_1} e^{-in\theta_0}}{\sqrt{2\pi|\lambda_1|}} \frac{e^{i\lambda_1 \rho}}{\sqrt{\rho}}
\left[ \begin{array}{l} a \\ a\ e^{i\theta} \\ b\ e^{-i\theta} \\ b \end{array} \right] \right. \nonumber \\
&& + \left. (-1)^n \frac{ e^{i\frac{\pi}{4}\gamma_1} e^{-in\theta_0}}{\sqrt{2\pi|\lambda_1|}} \frac{e^{-i\lambda_1 \rho}}{\sqrt{\rho}}
\left[ \begin{array}{l} a \\ -a\ e^{i\theta} \\ b\ e^{-i\theta} \\ -b \end{array} \right] \right\} \nonumber \\
\label{eq:PhiPLPW}
\end{eqnarray}
and the hole-like plane wave takes the form
\begin{eqnarray}
\Phi_{PW}^h(\rho,\theta) && \!\!\!\!\! = \! \sum_{n=-\infty}^\infty e^{in\theta} \left\{ \frac{ e^{-i\frac{\pi}{4}\gamma_2} e^{-in\theta_0}}{\sqrt{2\pi|\lambda_2|}} \frac{e^{i\lambda_2 \rho}}{\sqrt{\rho}}
\left[ \begin{array}{l} b \\ -b\ e^{i\theta} \\ a\ e^{-i\theta} \\ -a \end{array} \right] \right. \nonumber \\
&& + \left. (-1)^n \frac{ e^{i\frac{\pi}{4}\gamma_2} e^{-in\theta_0}}{\sqrt{2\pi|\lambda_2|}} \frac{e^{-i\lambda_2 \rho}}{\sqrt{\rho}}
\left[ \begin{array}{l} b \\ b\ e^{i\theta} \\ a\ e^{-i\theta} \\ a \end{array} \right] \right\} . \nonumber \\
\label{eq:PhiHLPW}
\end{eqnarray}
Note that these expressions now bear a striking resemblance to our general expression for the asymptotic wave function, Eq.~(\ref{eq:PhiAsymptotic}), with the particle-like plane wave contributing only to the particle-like terms (1 and 3) and the hole-like plane wave contributing only to the hole-like terms (2 and 4).

\subsection{Outgoing radial waves}
\label{ssec:outgoing}
Now we need only subtract each plane wave from the total asymptotic wave function and identify what remains as scattered waves.
\begin{equation}
\Phi(\rho,\theta) = \Phi_{PW}(\rho,\theta) + \Delta\Phi(\rho,\theta)
\label{eq:DeltaPhidef}
\end{equation}
Let's begin with the case of an incident particle-like plane wave.  Subtracting Eq.~(\ref{eq:PhiPLPW}) from Eq.~(\ref{eq:PhiAsymptotic}) reveals four possible scattered radial waves
\begin{eqnarray}
\Delta\Phi^p &=&
F_{11} \frac{e^{ik_1 r}}{\sqrt{r}} \left[ \begin{array}{l} a \\ a\ e^{i\theta} \\ b\ e^{-i\theta} \\ b \end{array} \right]
+ F_{12} \frac{e^{ik_2 r}}{\sqrt{r}} \left[ \begin{array}{l} b \\ -b\ e^{i\theta} \\ a\ e^{-i\theta} \\ -a \end{array} \right] \nonumber \\
&+& F_{13} \frac{e^{-ik_1 r}}{\sqrt{r}} \left[ \begin{array}{l} a \\ -a\ e^{i\theta} \\ b\ e^{-i\theta} \\ -b \end{array} \right]
+ F_{14} \frac{e^{-ik_2 r}}{\sqrt{r}} \left[ \begin{array}{l} b \\ b\ e^{i\theta} \\ a\ e^{-i\theta} \\ a \end{array} \right] \nonumber \\
\label{eq:DeltaPhiPL}
\end{eqnarray}
with scattering amplitudes
\begin{equation}
F_{1j}(\varphi) \equiv \tfrac{1}{\sqrt{k_0}} \sum_n s_n^{1j} e^{in\varphi}
\label{eq:F1j}
\end{equation}
where $j=\{1,2,3,4\}$, $\varphi \equiv \theta-\theta_0$, and
\begin{eqnarray}
s_n^{11} &\equiv& C_n^1 e^{in\theta_0} - \frac{e^{-i\frac{\pi}{4} \gamma_1}}{\sqrt{2\pi |\lambda_1|}}
\;\;\;\;\;\; s_n^{12} \equiv C_n^2 e^{in\theta_0} \nonumber \\
s_n^{13} &\equiv& C_n^3 e^{in\theta_0} - (-1)^n \frac{e^{i\frac{\pi}{4} \gamma_1}}{\sqrt{2\pi |\lambda_1|}}
\;\;\;\;\;\; s_n^{14} \equiv C_n^4 e^{in\theta_0} .
\label{eq:s1s2s3s4PL}
\end{eqnarray}
But if we consider the current due to each of these terms, via Eq.~(\ref{eq:jqpFromComponents}),
\begin{eqnarray}
{\bf j}_{qp}[\Phi_{S1}^p] &=& 2 (a^2 - b^2) v \frac{|F_{11}(\varphi)|^2}{r} \hat{\bf r} \nonumber \\
{\bf j}_{qp}[\Phi_{S2}^p] &=& 2 (a^2 - b^2) v \frac{|F_{12}(\varphi)|^2}{r} \hat{\bf r} \nonumber \\
{\bf j}_{qp}[\Phi_{S3}^p] &=& -2 (a^2 - b^2) v \frac{|F_{13}(\varphi)|^2}{r} \hat{\bf r} \nonumber \\
{\bf j}_{qp}[\Phi_{S4}^p] &=& -2 (a^2 - b^2) v \frac{|F_{14}(\varphi)|^2}{r} \hat{\bf r} \nonumber \\
\label{eq:4currents}
\end{eqnarray}
and recall that $a^2-b^2>0$, it is clear that only the first two terms, with current in the $\hat{\bf r}$ direction, yield outgoing radial waves.  The last two terms, with current in the $-\hat{\bf r}$ direction, describe incoming radial waves, which must be set to zero.  Doing so imposes our asymptotic boundary condition and requires that $s_n^{13} = s_n^{14} = 0$ for all $n$, which specifies two of our previously-undetermined coefficients
\begin{equation}
C_n^3 = (-1)^n e^{-in\theta_0} \frac{e^{i\frac{\pi}{4} \gamma_1}}{\sqrt{2\pi |\lambda_1|}}
\;\;\;\;\;\;
C_n^4 = 0 .
\label{eq:Cn3Cn4PL}
\end{equation}
Thus, for the case of an incident particle-like plane wave, the full wave function is
\begin{eqnarray}
\lefteqn{\Phi^p(\rho,\theta) = e^{i{\bf k}_1 \cdot {\bf r}}
\left[ \begin{array}{l} a \\ a\ e^{i\theta_0} \\ b\ e^{-i\theta} \\ b\ e^{-i\theta} e^{i\theta_0} \end{array} \right]} \\
&& + F_{11}(\varphi) \frac{e^{ik_1 r}}{\sqrt{r}} \left[ \begin{array}{l} a \\ a\ e^{i\theta} \\ b\ e^{-i\theta} \\ b \end{array} \right]
+ F_{12}(\varphi) \frac{e^{ik_2 r}}{\sqrt{r}} \left[ \begin{array}{l} b \\ -b\ e^{i\theta} \\ a\ e^{-i\theta} \\ -a \end{array} \right] \nonumber
\label{eq:PhiPL}
\end{eqnarray}
where $F_{11}(\varphi)$ is the amplitude for same-channel scattering while $F_{12}(\varphi)$ is the amplitude for cross-channel scattering.

For the case of an incident hole-like plane wave, the analogous calculation proceeds just as above, except now it is the second and fourth terms that get a contribution from the plane wave.  The resulting wave function takes the form
\begin{eqnarray}
\lefteqn{\Phi^h(\rho,\theta) = e^{i{\bf k}_2 \cdot {\bf r}}
\left[ \begin{array}{l} b \\ -b\ e^{i\theta_0} \\ a\ e^{-i\theta} \\ -a\ e^{-i\theta} e^{i\theta_0} \end{array} \right]} \\
&& + F_{21}(\varphi) \frac{e^{ik_1 r}}{\sqrt{r}} \left[ \begin{array}{l} a \\ a\ e^{i\theta} \\ b\ e^{-i\theta} \\ b \end{array} \right]
+ F_{22}(\varphi) \frac{e^{ik_2 r}}{\sqrt{r}} \left[ \begin{array}{l} b \\ -b\ e^{i\theta} \\ a\ e^{-i\theta} \\ -a \end{array} \right] \nonumber
\label{eq:PhiHL}
\end{eqnarray}
where
\begin{equation}
F_{2j}(\varphi) \equiv \tfrac{1}{\sqrt{k_0}} \sum_n s_n^{2j} e^{in\varphi}
\label{eq:F2j}
\end{equation}
\begin{equation}
s_n^{21} \equiv C_n^1 e^{in\theta_0}
\;\;\;\;\;\;
s_n^{22} \equiv C_n^2 e^{in\theta_0} - \frac{e^{-i\frac{\pi}{4} \gamma_2}}{\sqrt{2\pi |\lambda_2|}} .
\label{eq:t1t2HL}
\end{equation}
And in this case, the no-incoming-radial waves condition requires that
\begin{equation}
C_n^3 = 0
\;\;\;\;\;\;
C_n^4 = (-1)^n e^{-in\theta_0} \frac{e^{i\frac{\pi}{4} \gamma_2}}{\sqrt{2\pi |\lambda_2|}} .
\label{eq:Cn3Cn4HL}
\end{equation}
Note that the scattering amplitudes defined herein, $F_{11}$, $F_{12}$, $F_{21}$, and $F_{22}$, depend on the still-undetermined complex constants $C_n^1$ and $C_n^2$.  Thus, our next step is to determine these constants (and all others) by imposing boundary conditions.

\subsection{Matching solutions at boundaries}
\label{ssec:matching}
The solutions to the radial equations obtained in Sec.~\ref{sec:radial} were determined as a function of four complex constants per angular momentum index $n$, in each of three regions of our two-dimensional space ($0 < \rho < k_0\xi$, $k_0\xi < \rho < k_0\lambda$, and $k_0\lambda < \rho < \infty$) for a total of twelve constants per $n$.  As defined in Eqs.~(\ref{eq:xn}), (\ref{eq:yn}), and (\ref{eq:zn}), these are $A_n^j$, $B_n^j$, and $C_n^j$, where $j$ runs from 1 to 4.  They are specified via twelve constraints (per $n$): square-integrability as $\rho\rightarrow 0$ (2 constraints), continuity of four-vector solution at $\rho=k_0\xi$ (4 constraints), continuity of four-vector solution at $\rho=k_0\lambda$ (4 constraints), and no incoming radial waves as $\rho\rightarrow\infty$ (2 constraints).  Of these, the first two have already been used to specify $A_n^2$ and $A_n^4$ as per Eq.~(\ref{eq:A2A4zero}), and the last two have already been used to specify $C_n^3$ and $C_n^4$ as per Eqs.~(\ref{eq:Cn3Cn4PL}) and (\ref{eq:Cn3Cn4HL}).  This leaves eight constants (per $n$) to be determined by the eight continuity constraints
\begin{equation}
{\bf x}_n(k_0\xi) = {\bf y}_n(k_0\xi)
\;\;\;\;\;\;
{\bf y}_n(k_0\lambda) = {\bf z}_n(k_0\lambda)
\label{eq:continuityconstraints}
\end{equation}
which are easily recast in the form of the following $8 \times 8$ matrix equation
\begin{equation}
\left[ \begin{array}{cccccccc}
 & & & & & & 0 & 0 \\
{\bf x}_\xi^1 & {\bf x}_\xi^2 & -{\bf y}_\xi^1 & -{\bf y}_\xi^2 & -{\bf y}_\xi^3 & -{\bf y}_\xi^4 & 0 & 0 \\
 & & & & & & 0 & 0 \\
 & & & & & & 0 & 0 \\
0 & 0 & & & & & & \\
0 & 0 & -{\bf y}_\lambda^1 & -{\bf y}_\lambda^2 & -{\bf y}_\lambda^3 & -{\bf y}_\lambda^4 & {\bf z}_\lambda^1 & {\bf z}_\lambda^2 \\
0 & 0 & & & & & & \\
0 & 0 & & & & & &
\end{array} \right]
\left[ \begin{array}{c} A_n^1 \\ A_n^3 \\ B_n^1 \\ B_n^2 \\ B_n^3 \\ B_n^4 \\ C_n^1 \\ C_n^2 \end{array} \right]
= \left[ \begin{array}{c} 0 \\ 0 \\ 0 \\ 0 \\ \\ {\bf q} \\ \\ \\ \end{array} \right]
\label{eq:linearsolve}
\end{equation}
where ${\bf q} \equiv -C_n^3 {\bf z}_n^3(k_0\lambda) - C_n^4 {\bf z}_n^4(k_0\lambda)$ and ${\bf x}_\xi^j$ is shorthand for the four-component column vector ${\bf x}_n^j(k_0\xi)$.  We solve the above numerically, for each $n$, to obtain the eight-vector of constants, including the $C_n^1$ and $C_n^2$ that determine the scattering amplitudes.

\subsection{Cross sections}
\label{ssec:crosssections}
With the scattering amplitudes defined and the constants determined, we can proceed to write down expressions for the differential, total, transport, and skew cross sections.  Neglecting the rapidly oscillating cross-terms between the two scattering channels (as per Ref.~\onlinecite{cle68}), we can define the differential cross section in terms of the currents associated with the incident plane wave and each of the outgoing radial waves:
\begin{equation}
\frac{d\sigma}{d\varphi} = \frac{{\bf j}_{qp}[\Phi_{S1}] \cdot {\bf r} + {\bf j}_{qp}[\Phi_{S2}] \cdot {\bf r}}{{\bf j}_{qp}[\Phi_{PW}] \cdot \hat{\bf k}} .
\label{eq:dcsdef}
\end{equation}
Plugging in the currents via Eqs.~(\ref{eq:currentPLPW}) and (\ref{eq:4currents}), and noting that common factors of $2(a^2-b^2)v$ cancel out, this reduces to
\begin{equation}
\frac{d\sigma}{d\varphi} = |F_{\ell 1}(\varphi)|^2 + |F_{\ell 2}(\varphi)|^2
\label{eq:dcs}
\end{equation}
\begin{equation}
F_{\ell j}(\varphi) = \tfrac{1}{\sqrt{k_0}} \sum_n s_n^{\ell j} e^{in\varphi}
\label{eq:Fij}
\end{equation}
where $\ell=1$ for the incident particle-like case and $\ell=2$ for the incident hole-like case.

Integrating over $\varphi=\theta-\theta_0$ gives the total cross section
\begin{equation}
\sigma \equiv \int_{-\pi}^\pi \frac{d\sigma}{d\varphi} d\varphi = \frac{2\pi}{k_0} \sum_{j=1}^2 \sum_n |s_n^{\ell j}|^2 .
\label{eq:cs}
\end{equation}
The transport cross section, defined with an extra factor of $1-\cos\varphi$ to weight transport-relevant backscattering more heavily than forward scattering, is then given by
\begin{eqnarray}
\sigma_\parallel &\equiv& \int_{-\pi}^\pi \frac{d\sigma}{d\varphi} (1-\cos\varphi) d\varphi \nonumber \\
&=& \frac{2\pi}{k_0} \sum_{j=1}^2 \sum_n s_n^{\ell j} [s_n^{\ell j} - (s_{n+1}^{\ell j} + s_{n-1}^{\ell j})/2]^* .
\label{eq:tcs}
\end{eqnarray}
And the skew cross section, weighted with a factor of $\sin\varphi$ and thereby sensitive to the skewness of the scattering, is obtained via
\begin{eqnarray}
\sigma_\perp &\equiv& \int_{-\pi}^\pi \frac{d\sigma}{d\varphi} \sin\varphi d\varphi \nonumber \\
&=& \frac{2\pi}{k_0} \sum_{j=1}^2 \sum_n s_n^{\ell j} [s_{n+1}^{\ell j} - s_{n-1}^{\ell j}]^* .
\label{eq:scs}
\end{eqnarray}
Our computations of all of the above, for relevant parameter values, are presented in Sec.~\ref{sec:results}.

\subsection{Divergent same-channel forward scattering}
\label{ssec:divergent}
One difficulty intrinsic to the single-vortex scattering problem, common to both the ordinary \cite{cle68} and topological cases, is that the same-channel scattering amplitude (and therefore differential cross section) diverges in the forward ($\varphi=0$) direction.  This is a direct consequence of the Aharonov-Bohm effect, identified in Aharonov and Bohm's original paper \cite{aha59}, and comes about because even in the asymptotic ($\rho\rightarrow\infty$) limit, the quasiparticle acquires a non-trivial Berry phase upon circling the vortex.  This aspect of the influence of the vortex is therefore infinite-ranged (within the single vortex model) and affects both the scattered wave and the incident plane wave, which cannot quite be separated in the forward direction, resulting in the divergence. \cite{aha59,ola85,ste95}

For us, the practical consequence is that, while the cross-channel scattering coefficients, $s_n^{j\bar{j}}$, vanish for large $|n|$, the same-channel coefficients, $s_n^{jj}$, do not, and therefore cannot all be computed numerically.  Fortunately, the same-channel coefficients do approach a constant value for large positive $n$ and a different constant value for large negative $n$.  Thus, while the total cross section is infinite, the form of Eqs.~(\ref{eq:tcs}) and  (\ref{eq:scs}) reveal that the transport and skew cross sections remain finite.  We can therefore obtain $\sigma_\parallel$ and $\sigma_\perp$ by running our numerics up to some $|n| = n_{\rm max}$ beyond which differences between nearby coefficients are negligible.

Furthermore, once these large-$|n|$ constants are known, we can break the same-channel scattering amplitude into two pieces:
\begin{equation}
F_{jj} = F_{jj}^D + F_{jj}^{ND} = \tfrac{1}{\sqrt{k_0}} \sum_n s_n^D e^{in\varphi} + \tfrac{1}{\sqrt{k_0}} \sum_n s_n^{ND} e^{in\varphi}
\label{eq:Fjjtwopieces}
\end{equation}
a divergent piece with an infinite number of simple terms that can be summed analytically, and a non-divergent piece with a finite number of terms that can be calculated and summed numerically.  Recall that the same-channel coefficients have the form
\begin{equation}
s_n^{jj} = C_n^j e^{in\theta_0} - \frac{e^{-i\frac{\pi}{4} \gamma_j}}{\sqrt{2\pi |\lambda_j|}} .
\label{eq:sjj}
\end{equation}
where $\gamma_j=\mbox{sgn}(\lambda_j)$ and the $C_n^j$ are calculated numerically for each $n$.  Our numerics reveal that, in all cases, the resulting $s_n^{jj}$ coefficients approach $i\gamma_j/\sqrt{\pi|\lambda_j|}$ for large positive $n$ and $-1/\sqrt{\pi|\lambda_j|}$ for large negative $n$.  Thus, if we define
\begin{equation}
s_n^D \equiv \frac{1}{\sqrt{\pi|\lambda_j|}} \left\{
\begin{array}{r} i\gamma_j \;\;\;\;\;\;\;\;\;\;\; {\rm for} \;\; n>0 \\
(-1 + i\gamma_j)/2  \;\; {\rm for} \;\; n=0 \\
-1 \;\;\;\;\;\;\;\;\;\;\;\; {\rm for} \;\; n<0 \end{array} \right.
\label{eq:sD}
\end{equation}
and subtract as follows
\begin{equation}
s_n^{ND} \equiv s_n^{jj} - s_n^D = C_n^j e^{in\theta_0} - \frac{e^{i\frac{\pi}{4} \gamma_j}}{\sqrt{2\pi |\lambda_j|}} \mbox{sgn}(n)
\label{eq:sND}
\end{equation}
then the resulting $s_n^{ND}$ coefficients will, by construction, vanish for large $|n|$, allowing us to numerically compute the non-divergent part of the scattering amplitude, $F_{jj}^{ND}$.  Then the divergent part, which does not depend on the $C_n^j$ coefficients, is obtained by calculating the resulting geometric series analytically:
\begin{eqnarray}
F_{jj}^{D} &=& \lim_{\epsilon \rightarrow 0} \tfrac{1}{\sqrt{k_0}} \sum_{n=-\infty}^\infty s_n^D e^{in\varphi} e^{-\epsilon |n|} \nonumber \\
&=& \frac{-e^{-i\frac{\pi}{4}\gamma_j}}{\sqrt{2\pi k_0 |\lambda_j|}}
\left[ 2\pi D(\varphi) + \gamma_j S(\varphi) \right]
\label{eq:FjjD}
\end{eqnarray}
where
\begin{equation}
D(\varphi) = \lim_{\epsilon \rightarrow 0} \frac{1}{2\pi} \frac{\sinh\epsilon}{\cosh\epsilon - \cos\varphi} = \delta(\varphi)
\label{eq:Ddef}
\end{equation}
\begin{equation}
S(\varphi) = \lim_{\epsilon \rightarrow 0} \frac{\sin\varphi}{\cosh\epsilon - \cos\varphi} = \frac{\sin\varphi}{1 - \cos\varphi} .
\label{eq:Sdef}
\end{equation}
Here we made use of a convergence factor, $e^{-\epsilon |n|}$, to sum the series and then took the $\epsilon \rightarrow 0$ limit.  Note that both $D(\varphi)$ (which reduces to a Dirac delta function) and $S(\varphi)$ diverge for $\varphi=0$.  Adding this divergent part to the non-divergent part yields a same-channel scattering amplitude, and via Eq.~(\ref{eq:dcs}), a differential cross section, that diverge in the forward direction, but can be computed nevertheless.  As we shall see in the following section, this divergence is weak enough that the transport and skew cross sections remain finite (except at special values of our input parameters).

\section{Results}
\label{sec:results}

\subsection{Parameter regimes}
\label{ssec:regimes}
\begin{figure}
\centerline{\resizebox{3.25in}{!}{\includegraphics{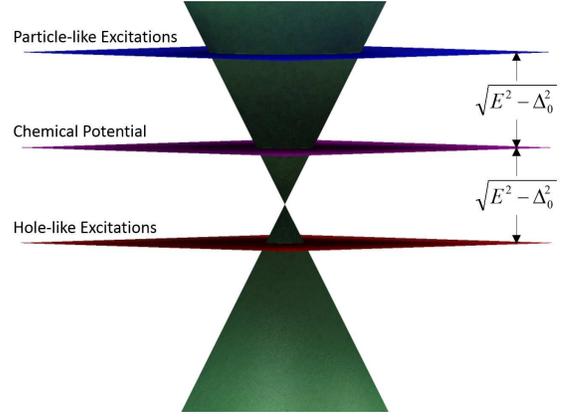}}}
\caption{(Color online) Depiction of the TI surface Dirac cone, $\varepsilon(k_x,k_y)$, intersected by three constant-$\varepsilon$ planes.  The central plane denotes the chemical potential $\mu$, while the top and bottom planes are shifted up and down from $\mu$ by $\sqrt{E^2-\Delta_0^2}$.  States at the intersection of the Dirac cone with the upper (lower) plane are the particle-like (hole-like) excitations of the proximity-induced superconductor.  Variation of the chemical potential shifts all three planes up and down, while variation of the quasiparticle excitation energy $E$ varies the separation between the planes.}
\label{fig:PHStatesOnCone}
\end{figure}

The algorithms developed in Secs.~\ref{sec:formulation}-\ref{sec:crosssection} were implemented numerically to compute the differential, transport, and skew cross sections for both particle-like and hole-like incident quasiparticles scattering from a single vortex.  Results were obtained as a function of quasiparticle excitation energy, $E$, and chemical potential, $\mu$, for coherence length and penetration depth values consistent with a typical type-II superconductor, $\xi=v/4\Delta_0$ and $\lambda=5\xi$.  To illustrate the role of $E$ and $\mu$ variation in this problem, we plot, in Fig.~\ref{fig:PHStatesOnCone}, the TI surface Dirac cone, $\varepsilon(k_x,k_y) = \pm v \sqrt{k_x^2+k_y^2}$, intersected by three planes.  The central plane denotes the chemical potential, and the top and bottom planes are shifted up and down from $\mu$ by $\sqrt{E^2-\Delta_0^2}$.  As per Eqs.~(\ref{eq:particlelike}) and (\ref{eq:holelike}), the particle-like ($+$) and hole-like ($-$) states satisfy $\varepsilon=\mu\pm\sqrt{E^2 - \Delta_0^2}$ and therefore reside at the intersection of the Dirac cone with the upper and lower planes, respectively.  Changing $\mu$ shifts the three planes up and down together, while changing $E$ brings them closer together and farther apart.  Since the states that participate in the scattering are those on the upper and lower circles of intersection, parameter variation of either type changes the radii of these circles and thereby modifies the scattering.

\begin{figure}
\centerline{\resizebox{3.25in}{!}{\includegraphics{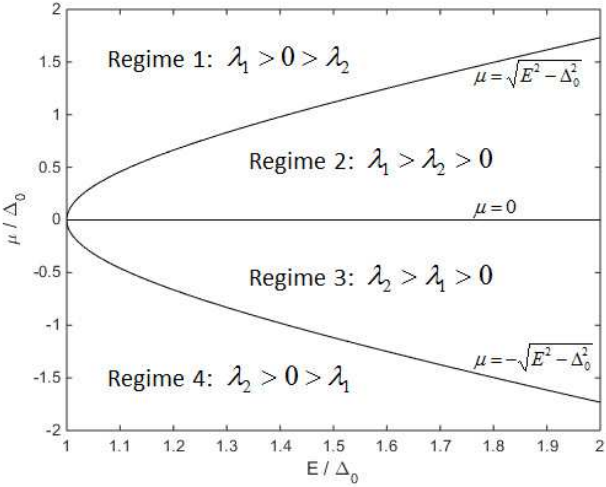}}}
\caption{Four parameter regimes in chemical potential ($\mu$) vs.\ quasiparticle excitation energy ($E$) space.  Note that at the interface between regime 1 and regime 2, eigenvalue $\lambda_2$ changes sign, while at the interface between regime 3 and regime 4, eigenvalue $\lambda_1$ changes sign.}
\label{fig:regimes}
\end{figure}

Of particular interest are the parameter values that pull either the particle-like or hole-like planes through the Dirac point, resulting in a coalescence of the circle of states to a single point at ${\bf k}=0$.  It is therefore convenient to define four parameter regimes based on which planes are above and below the Dirac point.  In regime 1 ($\mu > \sqrt{E^2-\Delta_0^2}$), all three planes are above the Dirac point.  In regime 2 ($0 < \mu < \sqrt{E^2-\Delta_0^2}$), the hole-like plane is below the Dirac point, with the other two above it.  (Note that the configuration depicted in Fig.~\ref{fig:PHStatesOnCone} corresponds to a point within regime 2.)  In regime 3 ($-\sqrt{E^2-\Delta_0^2} < \mu < 0$), the hole-like plane and the chemical potential are below the Dirac point, leaving only the particle-like plane above it.  And in regime 4 ($\mu < -\sqrt{E^2-\Delta_0^2}$), all three planes are below the Dirac point.  These four regimes are depicted in $\mu$ vs.\ $E$ parameter space in Fig.~\ref{fig:regimes}.  Note that we include only $E>\Delta_0$ in the parameter space.  For $E<\Delta_0$, our eigenvalues become complex and there are no oscillatory scattering states.  Note also that as the hole-like plane passes through the Dirac point, at the interface between regime 1 and regime 2, eigenvalue $\lambda_2$ passes through zero and changes sign.  Similarly, as the the particle-like plane passes through the Dirac point, at the interface between regime 3 and regime 4, eigenvalue $\lambda_1$ passes through zero and changes sign.

In what follows, we present numerical results at various points in the parameter space of Fig.~\ref{fig:regimes}.  We begin with the differential cross section at a single point in regime 1.  We then present transport and skew cross section results along two lines across the parameter space: a horizontal line at $\mu=\Delta_0$ that crosses from regime 1 to regime 2, and a vertical line at $E=\sqrt{2}\Delta_0$ that crosses all four regimes.

\subsection{Differential cross section}
\label{ssec:dCSresults}

\begin{figure}
\centerline{\resizebox{3in}{!}{\includegraphics{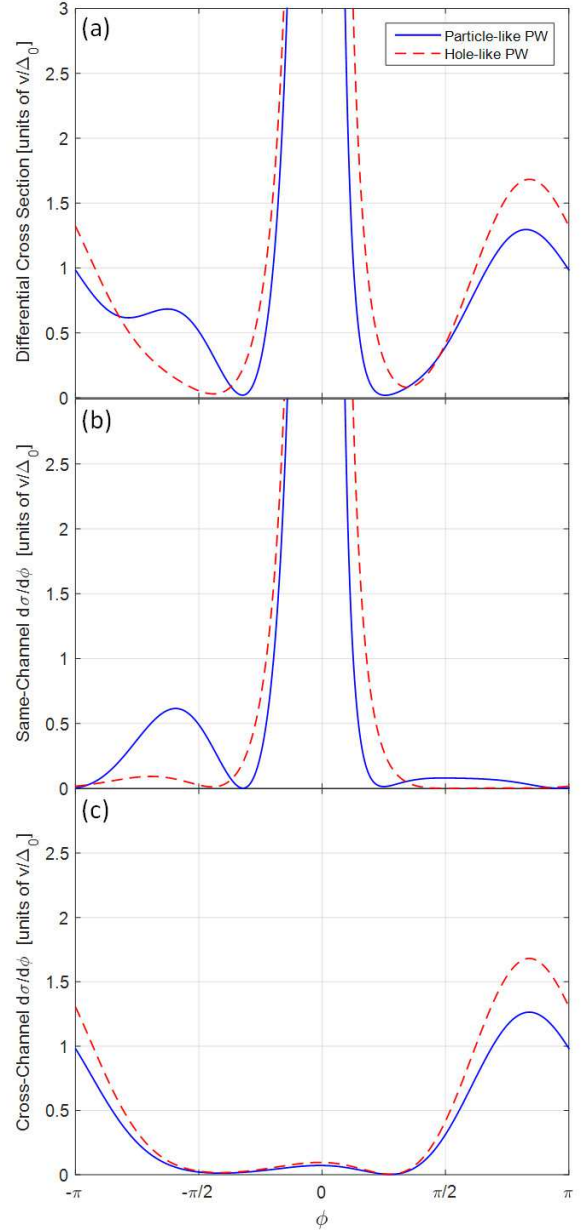}}}
\caption{(Color online) Calculated differential cross section, $d\sigma / d\varphi$, as a function of scattering angle $\varphi=\theta-\theta_0$, for $\mu=\Delta_0$, $E=1.01 \Delta_0$, $\xi = 0.25 v/\Delta_0$, and $\lambda = 5 \xi$.  Panel (a) shows the full differential cross section, while panels (b) and (c) show the same-channel and cross-channel contributions respectively.  In all panels, solid curves denote results for an incident particle-like plane wave while dashed curves denote results for an incident hole-like plane wave.}
\label{fig:dCS}
\end{figure}

As described in Sec.~\ref{ssec:divergent}, we obtain the differential cross section in three pieces: the cross-channel term (always non-divergent), the non-divergent part of the same-channel term, and the divergent part of the same-channel term.  The first two (non-divergent) pieces are computed numerically, including angular-momentum-indexed coefficients from $-n_{\rm max}$ to $n_{\rm max}$, beyond which additional coefficients are negligible.  Higher energies require more terms.  We included as many as 71 terms ($n_{\rm max} = 35$) in our computations.  The third (divergent in the forward direction) piece, obtained analytically via geometric series summation and given in Eqs.~(\ref{eq:FjjD})-(\ref{eq:Sdef}), is then included, and the differential cross section is computed via Eq.~(\ref{eq:dcs}) as a function of scattering angle, $\varphi = \theta-\theta_0$.  Presented in Fig.~\ref{fig:dCS}(a) are results for both particle-like and hole-like incident plane waves, for a particular point at the far left side of regime 1 in our parameter space ($\mu=\Delta_0$, $E=1.01 \Delta_0$).  As expected, results diverge in the forward ($\varphi=0$) direction because the vortex acts as an Aharonov-Bohm \cite{aha59} half flux.  Though the divergence in the forward direction yields an infinite total cross section via Eq.~(\ref{eq:cs}), it is sufficiently suppressed by the $1-\cos\varphi$ factor in Eq.~(\ref{eq:tcs}) as well as the $\sin\varphi$ factor in Eq.~(\ref{eq:scs}) that the resulting transport and skew cross sections are generally finite.

In panels (b) and (c), we separate out the same-channel and cross-channel contributions respectively.  Note that cross-channel scattering, while overwhelmed in the forward direction, is dominant in the back-scattering direction and therefore makes a significant contribution to the transport cross section.  This effect, most pronounced for $E$ just slightly greater than $\Delta_0$, is common to our topological case as well as the ordinary superconductor case described by Cleary \cite{cle68}.  It is a purely superconducting effect, due to the mixing of particle-like and hole-like excitations by the vortex.

\subsection{Transport cross section}
\label{ssec:tCSresults}

\begin{figure}
\centerline{\resizebox{3in}{!}{\includegraphics{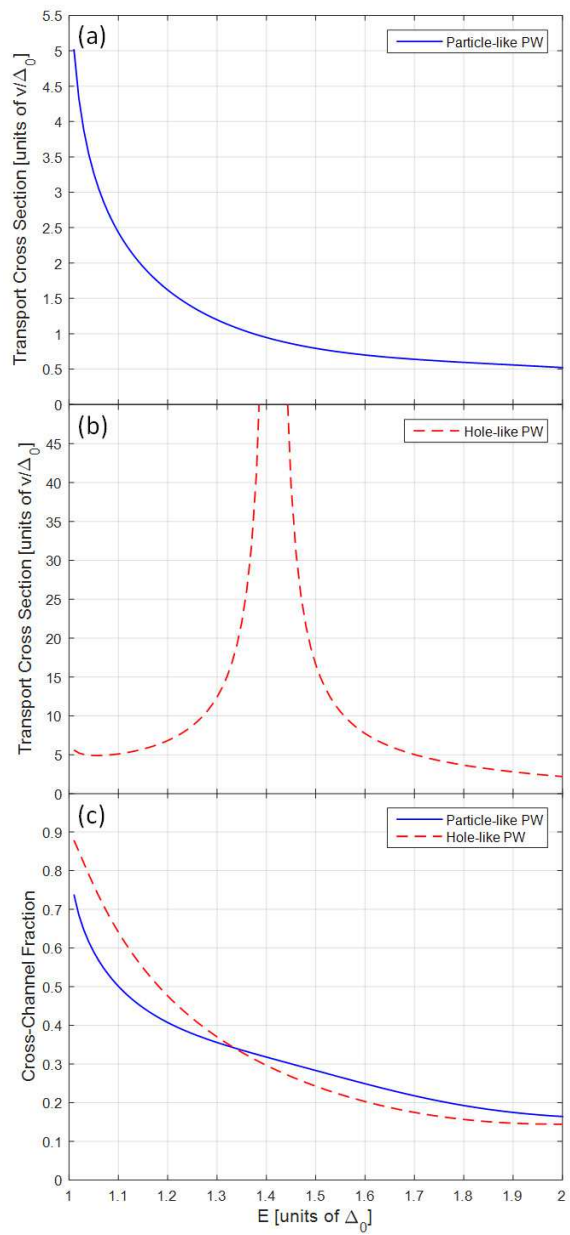}}}
\caption{(Color online) Calculated transport cross section, $\sigma_\parallel$, as a function of quasiparticle excitation energy, $E$, for $\mu=\Delta_0$, $\xi = 0.25 v/\Delta_0$, and $\lambda = 5 \xi$. Results for incident particle-like and hole-like plane waves are shown in panels (a) and (b) respectively.  For each case, the fraction of the transport cross section due to cross-channel scattering is shown in panel (c), plotted as a function of $E$.}
\label{fig:tCSvsE}
\end{figure}

Transport cross section, $\sigma_\parallel$, is computed via Eq.~(\ref{eq:tcs}) and we have done so for parameter values along two linear cuts through our parameter space.  Results for fixed chemical potential $\mu=\Delta_0$ as a function of quasiparticle excitation energy $E$ (a horizontal line across Fig.~\ref{fig:regimes} that brings us from regime 1 to regime 2) are plotted in Fig.~\ref{fig:tCSvsE}.  Panel (a) shows results for a particle-like plane wave, incident with $|{\bf k}|=k_0|\lambda_1|$.  The excitation energy dependence is quite similar to that observed for an ordinary superconductor \cite{cle68}, with $\sigma_\parallel$ falling off with increasing $E$ from a maximum value obtained at $E=\Delta_0$.  Since the particle-like plane remains above the Dirac point throughout both regime 1 and regime 2, $\lambda_1$ remains positive and the transition between the two regimes has little effect on the computed $\sigma_\parallel$.  Results are quite different for the case of a hole-like plane wave, incident with $|{\bf k}|=k_0|\lambda_2|$, shown in panel (b).  As $E$ surpasses $\sqrt{2}\Delta_0$, we transition from regime 1 to regime 2, the hole-like plane passes through the Dirac point, and the incident wave vector goes through zero as $\lambda_2$ changes sign.  The result, as seen in the plot, is a $\sigma_\parallel$ that diverges at $E=\sqrt{2}\Delta_0$ before decreasing as $E$ increases further.  (Of course, when we run the same calculation for negative $\mu$, the roles of particle-like and hole-like incident waves are reversed, with a divergence for the particle-like case and none for the hole-like case, as parameters are tuned from regime 3 to regime 4 and $\lambda_1$ (rather than $\lambda_2$) changes sign.)  This is an effect not seen in the ordinary superconductor case because there, the particle-like and hole-like planes are typically far from a band minimum.  By contrast, in the topological case, it is reasonable that the chemical potential could be tuned close enough to the Dirac point that any of our four regimes could be encountered.  Panel (c) shows the fraction of $\sigma_\parallel$ due to cross-channel scattering (either particle-like to hole-like or hole-like to particle-like).  In both cases, the cross-channel contribution is greatest for $E$ close to $\Delta_0$, where the separation between the particle-like and hole-like planes is minimal.  In this situation, the system is farthest from that of the normal state and the vortex is most effective in mixing particle-like and hole-like excitations, resulting in a greater cross section for excitations to change channels upon scattering.  Note that nothing remarkable happens to the cross-channel fraction at the $E=\sqrt{2}\Delta_0$ transition point, and that the fraction curves for both cases are quite similar, despite the divergence in $\sigma_\parallel$ for the incident hole-like case.

\begin{figure}
\centerline{\resizebox{3in}{!}{\includegraphics{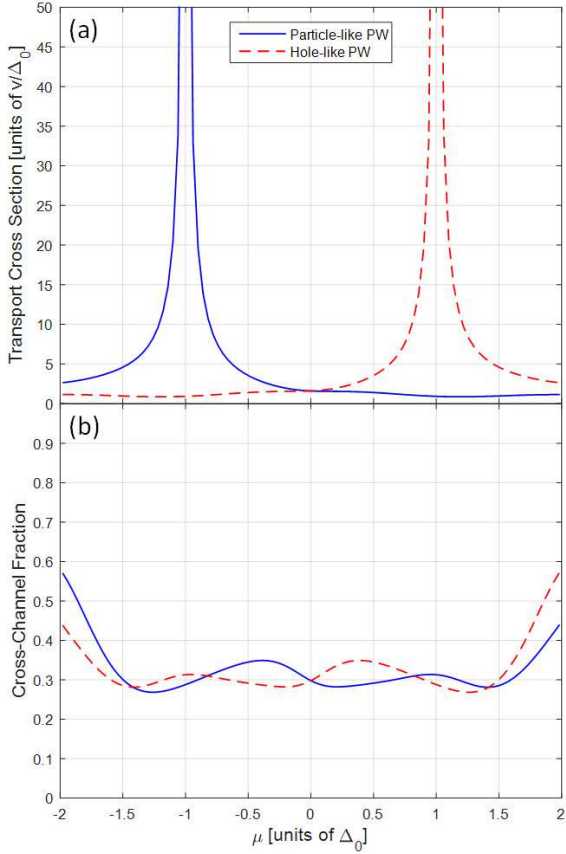}}}
\caption{(Color online) Calculated transport cross section, $\sigma_\parallel$, as a function of chemical potential, $\mu$, for $E=\sqrt{2}\Delta_0$, $\xi = 0.25 v/\Delta_0$, and $\lambda = 5 \xi$.  Results for incident particle-like (solid) and hole-like (dashed) plane waves are shown in panel (a).  The fraction of each due to cross-channel scattering is shown in panel (b), plotted as a function of $\mu$.}
\label{fig:tCSvsMU}
\end{figure}

Transport cross section results for fixed quasiparticle excitation energy $E=\sqrt{2}\Delta_0$ as a function of chemical potential $\mu$ (a vertical line through the parameter space of Fig.~\ref{fig:regimes}) are plotted in Fig.~\ref{fig:tCSvsMU}(a).  Here we cross all four parameter regimes: 4 to 3 to 2 to 1.  The transition from regime 4 to regime 3 sweeps the particle-like plane through the Dirac point, resulting in a divergence for the incident particle-like case at $\mu=-\Delta_0$.  Similarly, the transition from regime 2 to regime 1 sweeps the hole-like plane through the Dirac point, resulting in an equivalent divergence for the incident hole-like case at $\mu=\Delta_0$.  Panel (b) shows the cross-channel fraction as a function of $\mu$, which is roughly constant around 0.3 over most of the plot, increasing somewhat for larger $|\mu|$.  Note the particle-hole symmetry of these $\sigma_\parallel$ results, specifically that
\begin{equation}
\sigma_\parallel^h(\mu) = \sigma_\parallel^p(-\mu)
\label{eq:tCSphsymmetry}
\end{equation}
where the $p$ and $h$ superscripts denote the incident particle-like and incident hole-like cases respectively.  As we shall see in the following section, the situation is notably different for $\sigma_\perp$.

\subsection{Skew cross section}
\label{ssec:sCSresults}

\begin{figure}
\centerline{\resizebox{3in}{!}{\includegraphics{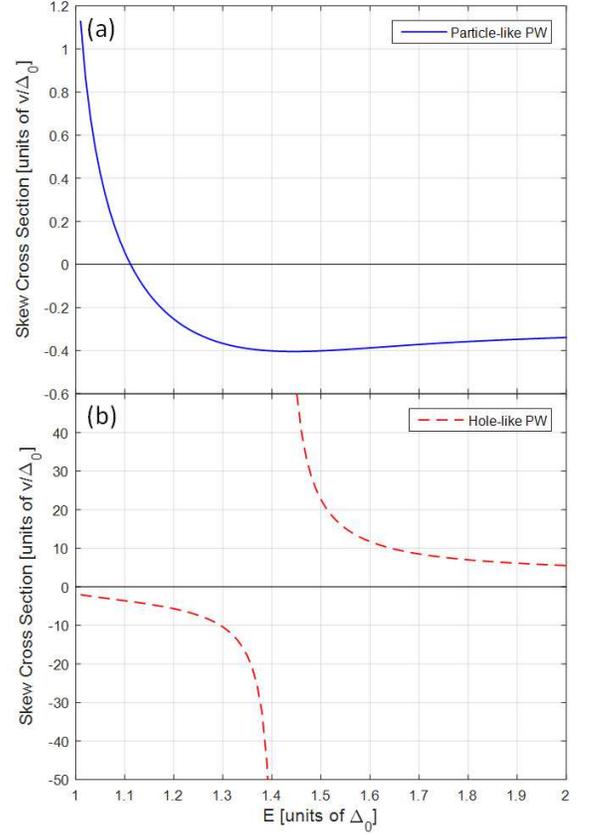}}}
\caption{(Color online) Calculated skew cross section, $\sigma_\perp$, as a function of quasiparticle excitation energy, $E$, for $\mu=\Delta_0$, $\xi = 0.25 v/\Delta_0$, and $\lambda = 5 \xi$. Results for incident particle-like and hole-like plane waves are shown in panels (a) and (b) respectively.}
\label{fig:sCSvsE}
\end{figure}

The skew cross section, $\sigma_\perp$, is computed via Eq.~(\ref{eq:scs}), and we present here results for parameter values along the same two traces through Fig.~\ref{fig:regimes} that we used previously in computing the transport cross section.  Excitation energy dependence for fixed chemical potential, $\mu=\Delta_0$, is plotted in Fig.~\ref{fig:sCSvsE}.  This trace takes us from regime 1 to regime 2, with the hole-like plane passing through the Dirac point at the transition energy, $E=\sqrt{2}\Delta_0$.  Panel (a) shows the incident particle-like case.  Here, the skew cross section is maximal, and positive, for $E$ close to $\Delta_0$.  As $E$ increases, $\sigma_\perp$ decreases, becomes negative, and reaches a negative minimum value before gradually increasing toward zero.  Note that the transition from regime 1 to regime 2 has little effect since the particle-like plane remains above the Dirac point throughout.  By contrast, in the incident hole-like case, shown in panel (b), the skew cross section diverges and changes sign upon transition from regime 1 to regime 2, as the hole-like plane passes through the Dirac point.

\begin{figure}
\centerline{\resizebox{3in}{!}{\includegraphics{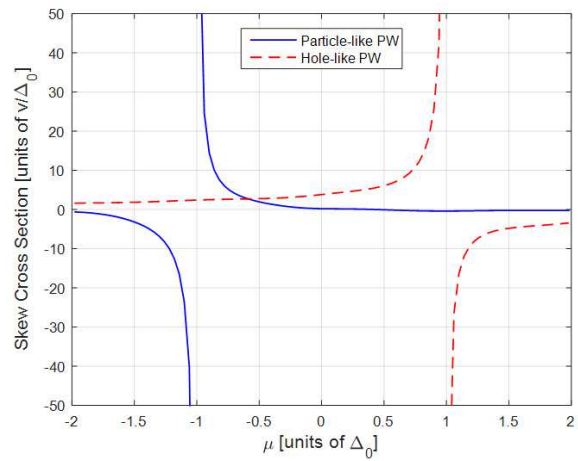}}}
\caption{(Color online) Calculated skew cross section, $\sigma_\perp$, as a function of chemical potential, $\mu$, for $E=\sqrt{2}\Delta_0$, $\xi = 0.25 v/\Delta_0$, and $\lambda = 5 \xi$.  The solid curve denotes results for an incident particle-like plane wave while the dashed curve denotes results for an incident hole-like plane wave.}
\label{fig:sCSvsMU}
\end{figure}

Chemical potential dependence for fixed quasiparticle excitation energy, $E=\sqrt{2}\Delta_0$, is plotted in Fig.~\ref{fig:sCSvsMU}.  This trace crosses all four regimes as $\mu$ is varied from negative values to positive.  Results resemble those we observed for $\sigma_\parallel$, with the skew cross section diverging as the appropriate planes pass through the Dirac point.  For an incident particle-like plane wave, $\sigma_\perp$ diverges and changes sign upon transition from regime 4 to regime 3, and for an incident hole-like plane wave , $\sigma_\perp$ diverges and changes sign upon transition from regime 2 to regime 1.  But in contrast to the particle-hole symmetry we observed in the transport cross section results, note here that $\sigma_\perp^h(\mu)\neq -\sigma_\perp^p(-\mu)$. Rather, we find that
\begin{equation}
\sigma_\perp^h(\mu) = -\sigma_\perp^p(-\mu) + \frac{4v}{\sqrt{E^2-\Delta_0^2}-\mu} .
\label{eq:sCSphnonsymmetry}
\end{equation}
This asymmetry is a consequence of the vortex, which breaks particle-hole symmetry in the problem.  The extra term above derives from the contribution to the skew cross section from the divergent part of the same-channel scattering amplitude, a result of the Aharonov-Bohm scattering of the vortex.  Only upon replacing the vortex by an antivortex is a skew cross section of the opposite sense obtained.

\section{Conclusions}
\label{sec:conclusions}
In this work, we have considered the scattering of quasiparticles from a vortex in the proximity-induced superconducting state at the 2D interface of a topological insulator and an $s$-wave superconductor.  This system is known to support a zero energy bound state at the vortex core, as was first shown by Fu and Kane \cite{fu08}.  Just as Cleary \cite{cle68} extended the Caroli-de Gennes-Matricon \cite{car64} bound state analysis of the vortex in an ordinary superconductor to include the scattering states, our purpose has been to study the scattering states of the vortex in the topological case.

Beginning with the Dirac-Bogoliubov-de Gennes equation, we separated variables in polar coordinates and established a hard-cutoff model that divides the system into three regions: inside the vortex core, inside the vortex but outside the core, and outside the vortex.  We then developed algorithms to solve the resulting radial equations in each of these three regions.  Applying boundary conditions between regions, as well as at the origin and at infinity, led to a definition of the scattering amplitudes and cross sections in terms of angular-momentum-indexed coefficients, as well as a procedure for solving for those coefficients.  As in the ordinary superconductor case, the vortex mixes particle-like and hole-like excitations.  Thus, ours is also a two channel problem, with incident particle-like plane waves scattering into outgoing particle-like and hole-like radial waves and incident hole-like plane waves scattering into outgoing hole-like and particle-like radial waves.  Also as in the ordinary superconductor case, the vortex acts as an Aharonov-Bohm half-flux, which leads to divergent same-channel scattering in the forward direction.  This results in an infinite total cross section but allows for the calculation of both the transport cross section, $\sigma_\parallel$, and the skew cross section, $\sigma_\perp$.

We calculated $\sigma_\parallel$ and $\sigma_\perp$ numerically, implementing the algorithms described above, as a function of both quasiparticle excitation energy, $E$, and chemical potential, $\mu$.  Results are best understood by considering the Dirac cone (i.e.\ as drawn in Fig.~\ref{fig:PHStatesOnCone}) and noting that the $k$-space location of particle-like excitations is given by the intersection of the Dirac cone with particle-like and hole-like planes shifted up and down, respectively, from the chemical potential by $\sqrt{E^2-\Delta_0^2}$.  When excitations of the incident channel (particle-like or hole-like) are far from the Dirac point, results are qualitatively similar to those obtained by Cleary \cite{cle68} for the case of an ordinary superconductor (see Fig.~\ref{fig:tCSvsE}(a)).  However, for parameter values where the incident plane passes through the Dirac point and the incident wave vector therefore approaches zero, the resulting transport and skew cross sections diverge as $1/|\lambda_j|$ and $1/\lambda_j$ respectively, where $\lambda_{1,2} \equiv (\sqrt{E^2-\Delta_0^2} \pm \mu)/\Delta_0$ (see Figs.~\ref{fig:tCSvsE}(b), \ref{fig:tCSvsMU}(a), \ref{fig:sCSvsE}(b), and \ref{fig:sCSvsMU}).  In an ordinary superconductor, analogous small-wave-vector quasiparticle states (for example, those deriving from a band minimum or maximum) are typically far from the chemical potential and therefore out of reach.  But in the topological case that we consider, it is presumed that the chemical potential can be tuned through the Dirac point, which is what brings the above effects into play.

We note, as Cleary \cite{cle68} did for ordinary superconductors, that the fraction of the transport cross section due to cross-channel scattering (particle-like to hole-like or vice versa) is maximal for quasiparticle excitation energies just above the superconducting gap, a consequence of the fact that cross-channel scattering, the mixing of particle-like and hole-like states by the vortex, is a purely superconducting effect.  We also note that this cross-channel fraction (see Figs.~\ref{fig:tCSvsE}(c) and \ref{fig:tCSvsMU}(b)) changes smoothly through the divergences in the cross sections, unaffected by the transit of particle-like and hole-like planes through the Dirac point.

Finally, we observe that taking $\mu$ to $-\mu$ exchanges incident particle-like transport cross section for incident hole-like transport cross section, $\sigma_\parallel^h(\mu) = \sigma_\parallel^p(-\mu)$.  However, the same particle-hole symmetry is not seen in the skew cross section, for which we find instead that $\sigma_\perp^h(\mu) = -\sigma_\perp^p(-\mu) + \frac{4v}{\sqrt{E^2-\Delta_0^2}-\mu}$ (see Figs.~\ref{fig:tCSvsMU} and \ref{fig:sCSvsMU}).  The extra term is contributed by the divergent part of the same-channel scattering (the Aharonov-Bohm part) and reflects the fact that the vortex breaks particle-hole symmetry in this problem.  A skew cross section of the opposite sense is only obtained by changing the direction of circulation of the vortex.

There are, of course, many directions in which this work could be extended in the future.  It would be instructive to repeat this analysis using a more detailed vortex model in place of the hard-cutoff model we introduced in Sec.~\ref{ssec:model}.  In addition, while our work has focused on the single vortex scattering problem and the calculation of the cross sections thereof, a logical next step is the application of these results to the calculation of transport coefficients in the presence of vortices, especially thermal transport coefficients since thermal current tracks quasiparticle current in this system.  Moving forward, it will be interesting to see if any of the divergences in the transport and skew cross sections yield peaks in the transport coefficients, or if they are suppressed by density-of-states factors that vanish as the particle-like and hole-like planes pass through the Dirac point.  Results will depend on a careful treatment of the interplay of vortex scattering and disorder.

\begin{acknowledgments}
I am grateful to B. Burrington, S. Ganeshan, and G. C. Levine for very helpful discussions.  This work was supported by faculty startup funds provided by Hofstra University.
\end{acknowledgments}

\end{document}